\documentclass[aps,prl,twocolumn,showpacs,superscriptaddress]{revtex4-1}

\usepackage{amsmath}
\usepackage{amsfonts}
\usepackage{amssymb}
\usepackage{amscd}
\usepackage{amsbsy}         
\usepackage{bm}             
\usepackage{graphicx}       
\usepackage{dcolumn}        
\usepackage{latexsym}
\usepackage[latin1]{inputenc}   
\usepackage{float}          

\newcommand  {\dip}   {\mathrm{dip}}
\newcommand  {\lat}   {\mathrm{lat}}
\newcommand  {\eff}   {\mathrm{eff}}

\newcommand  {\ev}    {\mathrm{ew}}

\newcommand  {\kB}    {k_\mathrm{B}}

\newcommand {\tub} {Physikalisches Institut, Eberhard-Karls-Universit\"{a}t T\"{u}bingen, Auf der Morgenstelle 14, D-72076 T\"{u}bingen, Germany}
\newcommand {\Brazil} {Institut de F\'isica de S\~ao Carlos, Universidade de S\~ao Paulo, 13560-970 S\~ao Carlos, SP, Brazil}

\begin{document}

\title{Photonic Band Gaps in One-Dimensionally Ordered Cold Atomic Vapors}

\author{Alexander Schilke}
\affiliation{\tub}
\author{Claus Zimmermann}
\affiliation{\tub}
\author{Philippe W. Courteille}
\affiliation{\Brazil}
\author{William Guerin}
\email{william.guerin@pit.uni-tuebingen.de}
\affiliation{\tub}
\date{\today}

\begin{abstract}
{We experimentally investigate the Bragg reflection of light at one-dimensionally ordered atomic structures by using cold atoms trapped in a laser standing wave. By a fine tuning of the periodicity, we reach the regime of multiple reflection due to the refractive index contrast between layers, yielding an unprecedented high reflectance efficiency of 80\%. This result is explained by the occurrence of a photonic band gap in such systems, in accordance with previous predictions.}
\end{abstract}

\pacs{37.10.Jk,42.25.Fx,42.70.Qs}

\maketitle


Cold atomic vapors can be used as suitable optical media for a number of applications or fundamental studies, in particular, in the fields of nonlinear and quantum optics, but also as complex optical media to study exotic wave transport phenomena, for instance coherent multiple scattering \cite{Labeyrie:2008}, weak localization \cite{Kaiser:2005} or random lasing \cite{Froufe:2009}. In these examples, an essential property is the \emph{disorder} inherent to atomic vapors that are simply confined in a magneto-optical trap (MOT).

On the contrary, \emph{ordered} clouds of cold atoms should exhibit different light-transport properties. The appearance of \emph{photonic band gaps} (PBGs) has indeed been predicted in one-dimensional lattices \cite{Deutsch:1995,Artoni:2005} and recently in three-dimensional, diamond (non-Bravais) lattices \cite{Antezza:2009}. The realization of PBGs in cold-atom samples would open the way to study new regimes of light transport in atomic vapors, where correlations and long-range order play a dominant role. The crossover regime, between order and disorder, or correlated disorder, is also a rich subject (see, e.g., \cite{Pasienski:2010}), in particular, in relation to Anderson localization \cite{Lagendijk:2009}.

\begin{figure*}[bt!]
\centering
\includegraphics{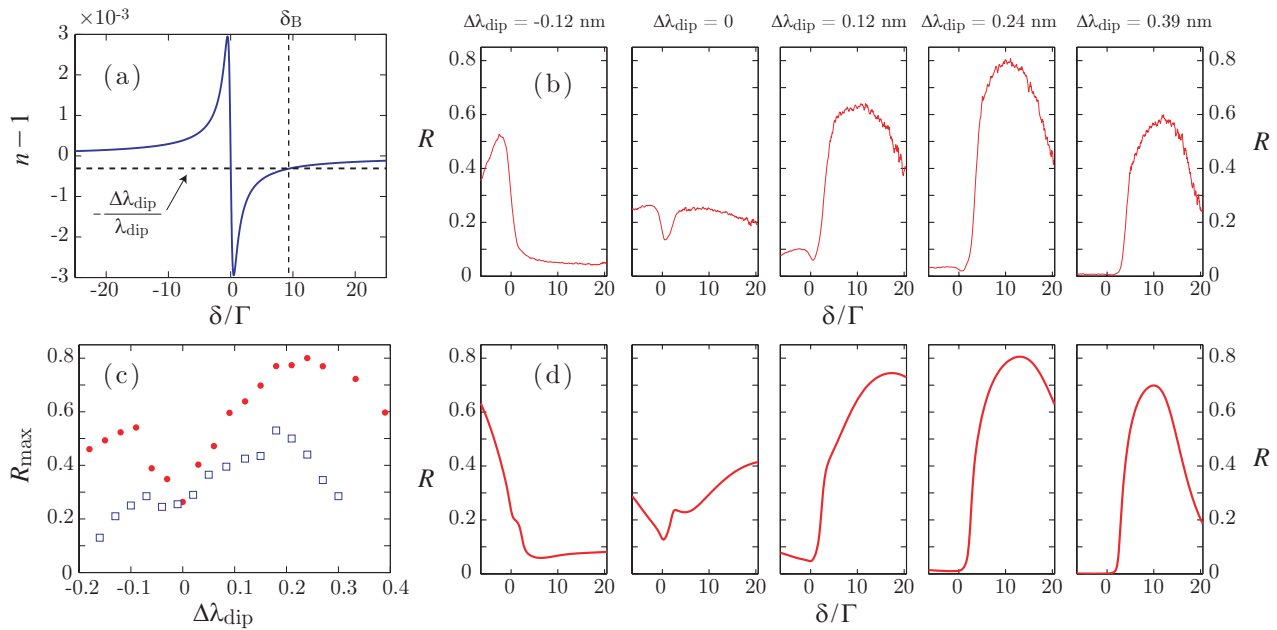}
\caption{(Color online). (a) Refractive index $n$ as a function of the normalized detuning $\delta/\Gamma$ for an averaged density $\rho = 7\times10^{11}$~cm$^{-3}$. The dashed lines represent the Bragg condition Eq.\ (\ref{eq.bragg_index}) with $\Delta\lambda_\dip = 0.24$~nm. (b) Experimental reflection spectra for different $\Delta\lambda_\dip$. (c) Measured maximal reflectance as a function of $\Delta\lambda_\dip$ for two different atom numbers, $N = 2\times 10^7$ (squares) and $N = 6 \times 10^7$ (dots). The spectra of (b) are taken from the latter. (d) Simulated reflection spectra.}
\label{fig.spectra}
\end{figure*}

However, creating efficient photonic structures is hard with cold atoms because these dilute systems have a low refractive index [Fig. \ref{fig.spectra}(a)] and a limited length. So far, efficient Bragg reflection of light has only been obtained with hot vapors, using an electromagnetically induced grating \cite{Bajcsy:2003}.
Previous investigations with cold atoms trapped in optical lattices have reported Bragg scattering with low efficiency. A first series of experiments used three-dimensional quasiresonant lattices \cite{Birkl:1995,Weidemuller:1995,Weidemuller:1998}, with a lattice geometry that does not create PBGs \cite{Antezza:2009}. Efficiencies below 1\% were reported. A second series of experiments investigated the 1D case \cite{Slama:2005a,Slama:2005b,Slama:2006}, where a band gap is expected, but the maximum reflectance was only 5\% in total power (30\% corrected from the partial overlap between the probe beam and the atomic sample) \cite{Slama:2006}. The limitations came mainly from the probing angle, which limited the interaction length with the lattice, and losses, i.e., out-of-axis scattering, due to the imaginary part of the atomic polarizability near resonance.

In this Letter, we report the observation of an efficient Bragg reflection at a one-dimensional lattice, reaching the regime of multiple reflections due to the refractive index contrast between layers and an 80\% efficiency in total power. This high efficiency is obtained by (1) using a small-diameter probe beam and a small probing angle to optimize the overlap with the atomic grating, and (2) adjusting the periodicity of the lattice so that the Bragg condition is fulfilled off the atomic resonance, thus strongly reducing the losses. Our experimental observations are explained by the appearance of a PBG, which we show to be robust against the system imperfections (finite length, varying density).



The experiment starts with a vapor-loaded MOT of $^{87}$Rb containing about $6\times10^8$ atoms. A dipole trap is generated by a homemade titanium-sapphire laser, whose available power is $1.3$~W and whose wavelength $\lambda_\dip$ is tunable. The beam is focussed on a waist ($1/e^2$ radius) $w_\dip = 220\,\mu$m at the MOT position (Rayleigh length $z_\mathrm{R} \simeq 0.2$~m). A 1D optical lattice is made by retroreflecting the beam, thus generating a structure whose periodicity is $\lambda_\dip/2$.

After stages of compression and molasses, the MOT is switched off and a waiting time of a few ms allows the untrapped atoms to fall down. Then, we can characterize the trapped sample with absorption imaging or acquire transmission and reflection spectra. Typical numbers for the trapped atoms are $N = 5\times 10^7$ atoms distributed over a length $L \sim 3$~mm ($\sim 7700$ layers). The temperature, $T \sim 100\, \mu$K, is related to the potential depth $U_0$ by a constant factor $\eta = U_0/\kB T \sim 3.5$. The transverse extension of the cloud is then $\sigma_\perp = w_\dip/(2\sqrt{\eta}) \simeq 60\, \mu$m  and the thickness of each layer along the lattice axis $z$ is $\sigma_z = \lambda_\dip/(2\pi\sqrt{2\eta}) \simeq 47$~nm (rms radii in the harmonic approximation).

To acquire spectra, we shine a weak and small (waist $w_0 = 35\,\mu$m), linearly polarized probe beam onto the lattice under an angle of incidence $\theta \simeq 2^\circ$. The angle is small enough so that the probe interacts with the lattice over its entire length and that the reflection is specular \cite{Slama:2005b}. The transmitted and reflected beams are then recorded with avalanche photodiodes. The probe frequency $\omega$ is swept in the vicinity of the atomic resonance $\omega_0$ ($F=2 \rightarrow F'=3$ closed transition of the D2 line, $\lambda_0 = 780.24$ nm, linewidth $\Gamma/2\pi=6.1$~MHz) by using an acousto-optical modulator in double-pass configuration. The other hyperfine levels are far enough to be negligible. The presented data are the result of an average of typically 100 cycles (the duration of each cycle is $\sim 1$~s).


Reflection occurs in the vicinity of a Bragg condition: the difference between the incident probe wavevector and the reflected wavevector must equal the lattice vector, i.e.\ $2 n(\delta) k_0 \cos \theta = K_\lat$, where $k_0=2\pi/\lambda_0$ is the probe wavevector in vacuum, $n$ is the real part of the \emph{average} refractive index of the medium, $\delta = \omega-\omega_0$ is the probe detuning from the atomic resonance, and $K_\lat = 4\pi/\lambda_\dip$ is the lattice vector \cite{footnote1}.

Experimentally, we keep the angle constant and adjust the Bragg condition by tuning the wavelength of the lattice beam. It is thus meaningful to rewrite the Bragg condition in the following form,
\begin{equation}\label{eq.bragg_index}
n(\delta)-1 = -\frac{\Delta\lambda_\dip}{\lambda_{\dip}} \; ,
\end{equation}
where $\Delta\lambda_\dip = \lambda_\dip- \lambda_{\dip0}$ is the shift from the ``geometric" (with $n=1$) Bragg condition $\lambda_{\dip0} = \lambda_0/ \cos \theta$.
With $\theta = 2^\circ$, $\lambda_{\dip0} \simeq 780.7$~nm. Then, for a given lattice wavelength, the Bragg condition is fulfilled for probe detunings $\delta$ given by Eq.\ (\ref{eq.bragg_index}), see Fig.\ \ref{fig.spectra}(a).
There are in general two such frequencies, but one is almost on resonance, where losses prevent any efficient reflection. The other Bragg frequency [$\delta_\mathrm{B}$ in Fig.\ \ref{fig.spectra}(a)] may be farther from resonance and can be tuned in order to search for an optimum.


Such an experiment is reported in Fig.\ \ref{fig.spectra}(b), which shows a set of spectra for different $\Delta\lambda_\dip$. As expected from Eq.\ (\ref{eq.bragg_index}), the spectra display a strong asymmetry, which evolves as the lattice wavelength is changed, the maximal efficiency going from one side of the resonance to the other side while $\Delta\lambda_\dip$ changes its sign. One can also clearly observe an optimum value of $\Delta\lambda_\dip$ for reaching high efficiencies, namely $R_\mathrm{max} \simeq 80 \%$ for $\Delta\lambda_\dip \simeq 0.24$~nm with our best atom number [Fig.\ \ref{fig.spectra}(c)]. Note that for each $\Delta\lambda_\dip$, we adjust the power accordingly to keep constant the potential depth and subsequently the atom number and temperature.

The existence of an optimum can easily be understood by considering the limiting cases. When $\Delta\lambda_\dip$ is very small, the Bragg condition is fulfilled where the refractive index is almost one, in virtue of Eq.\ (\ref{eq.bragg_index}), very far from resonance, leading to a very small index contrast in the periodic structure and thus to an inefficient reflection. In the opposite limit, when $\Delta\lambda_\dip$ is large, the Bragg condition is fulfilled near $\delta = \pm \Gamma/2$, where losses, due to the imaginary part of the atomic polarizability, are large and play a detrimental role. Ultimately, if $\Delta\lambda_\dip$ is too large for the given averaged atomic density $\rho$, the Bragg condition cannot be fulfilled at all. 
The optimum $\Delta\lambda_\dip$ depends on the average density $\rho$: a larger density allows us to increase $\Delta\lambda_\dip$, and thus the index contrast, without shifting the Bragg frequency towards resonance, i.e.\ without increasing losses. This is illustrated in Fig.\ \ref{fig.spectra}(c), which shows that the optimum $\Delta\lambda_\dip$ is larger with more atoms.

\begin{figure}[bt!]
\centering
\includegraphics{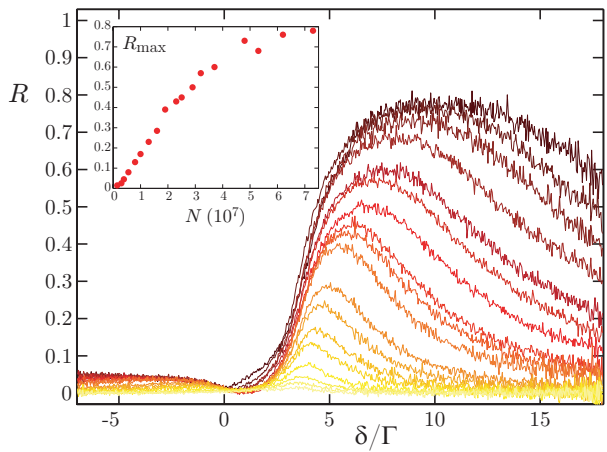}
\caption{(Color online). Spectra for different atom numbers $N$ in the lattice (constant length, varying density) with $\Delta\lambda_\dip = 0.15$~nm. Inset: Maximum reflectance as a function of $N$.}
\label{fig.spectrum_vsN}
\end{figure}

We have also studied the reflection spectrum as a function of the atom number for a given $\Delta\lambda_\dip$. In order to vary the atom number while keeping the lattice length constant, we have changed the waiting time in the lattice, before the spectra acquisition, from 5 ms to 400 ms, thus varying atom losses. The result is reported in Fig.\ \ref{fig.spectrum_vsN}. As expected from Eq.\ (\ref{eq.bragg_index}), we observe a shift farther from resonance together with a broadening for increasing densities. The evolution of the maximum reflectance as a function of the atom number (inset of Fig.\ \ref{fig.spectrum_vsN}) reveals that we almost reach saturation.

With such a high efficiency, it is natural to ask whether a photonic band gap occurs in our system and is responsible for the high reflectance. It is well-known that periodic 1D systems give rise to PBGs for any nonzero index modulation \cite{Joannopoulos}, so that no distinction is usually made between a 1D PBG and a Bragg reflector. However this is true only for infinite, lossless, and perfectly periodic media. As already noted, the atomic polarizability is complex so that our system has losses, and it is of course finite. Moreover, the lattice is not perfectly homogeneous but has a smooth density variation, so that the periodicity is not perfect. The situation is thus more intricate and deserves a careful analysis.


We can get a first insight with orders-of-magnitude estimations. A Bragg reflector made of repeated pairs of dielectric layers gives rise, in the stop band, to an evanescent wave whose penetration length (or Bragg length) is given by $L_\ev = \lambda_0/(4\Delta n)$, where $\Delta n$ is the refractive index difference between the two materials \cite{Brovelli:1995}.
By approximating the Gaussian atomic distribution in each well by a single layer of constant density with the same rms width and using Eq.\ (\ref{eq.bragg_index}), the index contrast is $\Delta n = \pi \sqrt{\eta/6} \times \Delta\lambda_\dip/\lambda_{\dip}$.
In our experiment, the maximum reflection is obtained for $\Delta\lambda_\dip \sim 0.24$~nm and we have $\eta \sim 3.5$, which gives a penetration length $L_\ev \sim 0.26$~mm, sensibly smaller than our optical lattice ($L\sim 3~$mm), so that we are indeed in the multiple-reflection regime. The effect of losses can also be evaluated by comparing the corresponding attenuation length $L_\mathrm{loss}$ to $L_\ev$. Given our estimated averaged density $\rho \sim 7\times 10^{11}$~cm$^{-3}$ and the detuning $\delta_\mathrm{B}$ determined by the Bragg condition (\ref{eq.bragg_index}), we estimate $L_\mathrm{loss}=1/\rho\sigma_\mathrm{sc}(\delta_\mathrm{B})$, where $\sigma_\mathrm{sc}(\delta_\mathrm{B})$ is the scattering cross section, which gives $L_\mathrm{loss} \sim 3.8$~mm, also much larger than $L_\ev$. Thus, this simple calculation confirms that our system fulfills the conditions $L_\ev \ll L, L_\mathrm{loss}$, which are necessary for the appearance of a band gap. A number of effects are however not taken into account, in particular the actual density distribution along the lattice, so that a more precise modeling is still valuable.


\begin{figure}[bt!]
\centering
\includegraphics{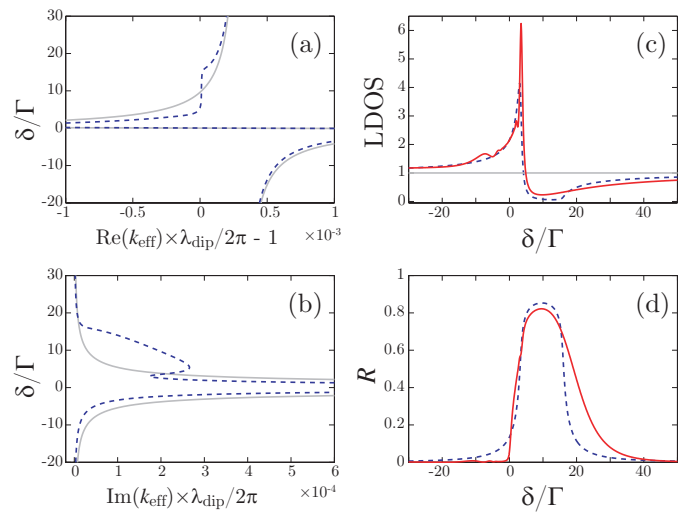}
\caption{(Color online). Modeling of our system. (a), (b) Real (a) and imaginary (b) parts of the effective wavevector $k_\eff$ in the medium as a function of the probe detuning $\delta$. The axes are reversed so that the graphs look like standard dispersion relations $\omega(k)$. (c) Normalized local density of states. (d) Reflection spectrum. In all plots, the light gray line corresponds to a homogeneous, nonperiodic medium of the same average density, the dashed line to an infinite lattice and the solid red (medium gray) line simulates the actual density distribution of our system.}
\label{fig.theory}
\end{figure}

We use the transfer matrix formalism to simulate the wave propagation in our system, and we refer to \cite{Deutsch:1995,Artoni:2005,Slama:2006} for detailed descriptions in similar contexts. From the transfer matrix of one single period, which takes into account the Gaussian atomic distribution in each well \cite{Slama:2006}, we draw the dispersion relations of the medium (effective wavevector $k_\eff$ vs $\omega$), valid in the limit of an infinitely long lattice (Bloch theorem). These are shown in Fig.\ \ref{fig.theory}(a,b), for the typical parameters of the experiment $\rho = 7\times10^{11}$~cm$^{-3}$, $\eta = 3.5$ and $\Delta\lambda_\dip = 0.24$~nm. The imaginary part of $k_\eff$ is composed of one Lorentzian of width $\Gamma$ centered on resonance, which corresponds to losses, and a supplementary part, leading to an evanescent wave, corresponding to a band gap [Fig.\ \ref{fig.theory}(b)]. In the same frequency range, the real part of $k_\eff$ has a reduced variation with $\omega$ [Fig.\ \ref{fig.theory}(a)], corresponding to a reduced density of states (DOS) $\mathcal{D} = d(\mathrm{Re}(k_\eff))/d\omega$. This last formula is however not always valid, especially in anomalous dispersive media, and we use the method of \cite{Boedecker:2003} to compute the normalized local DOS in the middle of the lattice. By using the complex reflection coefficients $r_1$, $r_2$ of the two surrounding, finite or infinite \cite{Deutsch:1995} semilattices, we obtain
\begin{equation}\label{eq.DOS}
\mathcal{D} = \mathrm{Re}\left[\frac{2+r_1+r_2}{1-r_1 r_2} -1\right] \; .
\end{equation}
The result, shown in Fig.\ \ref{fig.theory}(c), exhibits clearly the band gap, with a strong reduction of the normalized local DOS, which reaches at minimum 6\% for an infinite lattice. The corresponding reflection spectrum reaches 85\% [Fig. \ref{fig.theory}(d)]. In both cases, the limitation (for achieving high reflectance or low DOS) is the remaining losses.

To take into account the limited length of the sample and its actual, smooth density distribution along the lattice axis, that we characterized by absorption imaging, we compute the reflection spectra through the whole structure by multiplying elementary transfer matrices computed with the corresponding local density. The local DOS then exhibits a smaller reduction, reaching at minimum 23\%, but the maximum reflectance is almost as high as in the infinite limit, reaching 82\% [Fig. \ref{fig.theory}(c,d)].

Finally, to simulate our experimental spectra precisely, we must also take into account the trapping-induced inhomogeneous light shift and the finite transverse size. Indeed, we only considered so far infinitely extended layers, which is a crude approximation, the atom cloud having an rms width of $60\,\mu$m. The probe beam, having also a finite size, probes a distribution of local density, which induces an inhomogeneous broadening of the spectra. There is in fact a tradeoff for the probe beam size: a small beam allows us to probe a well-defined and maximum density, but since the reflection is very sensitive to the angle of incidence, the probe divergence induces also an inhomogeneous broadening. Experimentally, we have tried several probe sizes and obtained the maximum reflectance with $w_0 = 35\, \mu$m. To simulate these effects we averaged many spectra over Gaussian distributions of incident angles and densities corresponding, respectively, to the probe divergence and atomic density distribution along the transverse directions. We obtain the spectra of Fig.\ \ref{fig.spectra}(d), in good agreement with the experiment \cite{footnote2}.


To summarize, we have studied experimentally the Bragg reflection of light at a one-dimensional atomic quasiperiodic structure in the regime of multiple reflection, demonstrating a high efficiency, thanks to a fine tuning of the lattice periodicity. Then, motivated by the modeling of our system, we have investigated the effects of the finite length and of the smooth density variation along the lattice on the appearance of a band gap and its quality. These imperfections, inherent to cold-atom systems, do not destroy the band gap which, even if the DOS reduction is not dramatic, still substantially modifies the transport properties of light, as demonstrated by the observed high reflectance.

Improving further the quality of the PBG requires reducing the losses by creating the band gap farther from resonance, which could be done by increasing the density or the lattice length, for example with a larger initial MOT. Another fascinating possibility is to tailor the atomic polarizability to remove its absorptive part, while enhancing the refractive index, following the ideas of \cite{Scully:1991,Fleischhauer:1992}.
Finally, the extension to more evolved lattice geometries, for example with bichromatic lattices \cite{Rist:2009} and to three dimensions \cite{Antezza:2009}, has been proposed recently. This last case is of course the most appealing since a 3D band gap would profoundly modify the atom-light interaction \cite{Yablonovitch:1987,Lodahl:2004}.


We thank J.-F. Schaff and N. Mercadier for the computer-control program. We acknowledge support from the Alexander von Humboldt foundation, the DFG and the REA (program COSCALI, No. PIRSES-GA-2010-268717).



\begin{thebibliography}{26}
\bibitem[{foo({\natexlab{a}})}]{footnote1}
\bibinfo{note}{Note that contrary to the original situation considered in
  \cite{Deutsch:1995}, the refractive index for the lattice beam is
  negligible.}
\bibitem[{foo({\natexlab{b}})}]{footnote2}
\bibinfo{note}{The lower efficiency for $\Delta\lambda_\dip < 0$ might be due
  to heating induced by spontaneous emission, the dipole trap being closer to
  resonance.}
\end{thebibliography}


\end{document}